# Unconventional interlayer exchange coupling via chiral phonons in synthetic magnetic oxide heterostructures

**Authors**

Seung Gyo Jeong[1], Jiwoong Kim[2], Ambrose Seo[3], Sungkyun Park[2], Hu Young Jeong[4], Young-Min Kim[5], Valeria Lauter[6], Takeshi Egami[7,8], Jung Hoon Han[1], and Woo Seok Choi[1,*].

**Affiliations**

[1]Department of Physics, Sungkyunkwan University, Suwon 16419, Korea
[2]Department of Physics, Pusan National University, Busan 46241, Korea
[3]Department of Physics and Astronomy, University of Kentucky, Lexington, KY 40506, USA
[4]UNIST Central Research Facilities and School of Materials Science and Engineering, Ulsan National Institute of Science and Technology, Ulsan 44919, Korea
[5]Department of Energy Sciences, Sungkyunkwan University, Suwon 16419, Korea
[6]Neutron Scattering Division, Oak Ridge National Laboratory, Oak Ridge, TN 37831, USA
[7]Materials Science and Technology Division, Oak Ridge National Laboratory, Oak Ridge, TN 37831, USA
[8]Department of Materials Science and Engineering, University of Tennessee, Knoxville, TN 37996, USA
*Corresponding author. Email: choiws@skku.edu

**Abstract**

Chiral symmetry breaking of phonons plays an essential role in emergent quantum phenomena owing to its strong coupling to spin degree of freedom. However, direct experimental evidence of the chiral phonon-spin coupling is lacking. In this study, we report a chiral phonon-mediated interlayer exchange interaction in atomically controlled ferromagnetic metal ($SrRuO_3$)-nonmagnetic insulator ($SrTiO_3$) heterostructures. Owing to the unconventional interlayer exchange interaction, we have observed rotation of magnetic moments as a function of nonmagnetic insulating spacer thickness, resulting in a spin spiral state. The chiral phonon-spin coupling is further confirmed by phonon Zeeman effects. The existence of the chiral phonons and their interplay with spins along with our atomic-scale heterostructure approach open a window to unveil the crucial roles of chiral phonons in magnetic materials.

**Teaser**

Chiral phonons and their strong coupling with spins manifest unconventional interlayer exchange interaction and resultant novel spin state.

# MAIN TEXT

## Introduction

Chiral phonon serves as a fundamental element in realizing non-trivial quantum mechanical phenomena. When chiral symmetry breaks in a crystal, chiral phonons emerge and can couple to spins leading to phenomena such as phonon Hall effect, optically driven effective magnetic field, AC Stark effect, topologically-induced viscosity split, and pseudogap phase, as summarized in Table 1 (*1-16*). So far, thermal Hall and time-resolved spectroscopic measurements were employed for identifying the dynamic signatures of chiral phonons (*6-8*). However, a static manifestation of chiral phonons determining a ground state with long-range spin ordering is lacking.

Synthetic magnetic heterostructures let us study an interlayer exchange coupling (IEC), which might originate from the chiral phonon-spin coupling, e.g., the spin polarization of hole carriers by elliptically polarized phonons (*2*). In general, when two ferromagnetic (FM) layers are separated by a thin nonmagnetic-*metallic* (NM-*M*) spacer, the relative spin orientation of the FM layers is determined by the well-known Ruderman-Kittel-Kasuya-Yosida (RKKY) interaction via itinerant electrons (Fig. 1A). Here, the interlayer exchange constant ($J$) between the localized spins in the FM layers oscillates as a function of the spacer layer thickness ($t$) with the term $\cos(2k_F t) / t^d$, where $k_F$ and $d$ are the Fermi wavevector of the itinerant electrons in the spacer and the dimensionality of the system, respectively (*17, 18*). Hence, both parallel and antiparallel spin ordering between the FM layers can be realized depending on $t$, where the antiparallel alignment is commonly referred to as synthetic antiferromagnetic (sAFM) ordering. In contrast, for a nonmagnetic-*insulating* (NM-*I*) spacer without any itinerant carriers, $J$ is known to decay monotonically and exponentially with $t$, above the quantum tunneling thickness regime (*18, 19*). In the presence of chiral phonons, however, we propose that spins in the FM layers can indirectly interact with each other through a NM-I spacer via chiral phonon-spin coupling. Furthermore, depending on $t$, the interaction effectively changes the relative spin orientations of each FM layer (Figs. 1B and 1C).

We employed atomically designed $SrRuO_3$/$SrTiO_3$ (SRO/STO) superlattices to model the FM-M / NM-I / FM-M heterostructure. FM SRO (FM-M layer) is an excellent candidate for the realization of unconventional IEC, owing to its strong spin-phonon and spin-orbit coupling (*20-22*). A small Fermi energy mismatch between SRO and STO (NM-I layer) and their non-polar nature highly suppresses the charge transfer across the interface (*23-25*). Furthermore, identical *A*-site ion (*i.e.*, Sr) and similar in-plane lattice constants between SRO and STO provide coherent superlattice structures with fully strained states, *i.e.*, without misfit dislocations, of which the configuration amplifies the experimental signal (*21, 24*). In particular, the ferromagnetic molecular field of SRO layers breaks the degeneracy, enabling disparate population of the chiral phonons. The structural similarity of the perovskites and atomically sharp interfaces further facilitate the chiral phonon propagation, allowing chiral phonon-spin interaction in the neighboring SRO layers.

We deliberately grew oxide superlattices with alternating $α$ and $β$ atomic unit cells (~0.4 nm) of SRO ($t_{SRO}$) and STO ($t_{STO}$), respectively, repeated for $γ$ times on STO (001) substrates, *i.e.*, $[\alpha|\beta]_\gamma$, using pulsed laser epitaxy (See methods and Figs. S1-S3) (*21, 24-27*). An intriguing oscillation in the in-plane magnetization is observed as a function of $t_{STO}$, indicating the presence of an unconventional IEC across the NM-I spacer. The non-collinear spiral spin state in the ground state responsible for the observed magnetic oscillation was visualized by polarized neutron reflectometry (PNR). We propose chiral phonon-spin coupling as a possible mechanism for the unconventional IEC across the NM-I spacer. In particular, we suggest that the angular momentum of the chiral phonons couple to the spins, leading to the spin spiral state in our superlattices. Strong spin-phonon coupling was indeed evidenced by confocal Raman spectroscopy. This new type of long-range magnetic interaction yields an advanced understanding of the chiral phonons and accessible controllability of spiral spin states over them via atomic-scale heterostructuring. The spatial modulation of in-plane magnetization might further provide insight into the emergence of exotic spin orderings such as magnetic cone or fan (*28, 29*).

## Results

### NM-I thickness-dependent oscillatory magnetization behavior

The SRO/STO superlattices exhibit an exotic in-plane magnetic behavior that has not been observed in single SRO films (Figs. 1D-1G and Fig. S4). Temperature- (*T*-) dependent magnetization (*M* (*T*)) curves of the $[6|\beta]_{10}$ superlattices along the in-plane direction, for example, commonly reveal a typical FM $T_c$ = ~130 K, followed by an anomalous Néel-like downturn at ~80 K (Fig. 1D). Note that the out-of-plane *M* (*T*) shows the conventional FM behaviors of SRO with the same $T_c$ and without any downturn, consistent with previous studies (Fig. S5) (*21, 24-27*). Magnetic field- (*H*-) dependent in-plane magnetization (*M* (*H*)) curves of the superlattices consistently support the AFM-like behavior, with a double hysteresis loop and a large coercive field of ~1.8 T appearing below 80 K (Figs. 1E and 1F, Fig. S6). Furthermore, *H*-dependent *M* (*T*) curves show a gradual disappearance of the downturn as *H* is increased, especially above coercive field (Fig. S6). Similar behavior was reported in low-dimensional materials and understood as the AFM-like IEC between the FM layers (*19, 30*).

A clear $t_{STO}$-dependent oscillation of in-plane magnetization was observed, demonstrating the presence of unconventional IEC. We extracted *M* (2 K, 0.01 T) values as a function of both $t_{SRO}$ and $t_{STO}$, which exhibit a distinct oscillation as a function of $t_{STO}$ when $t_{SRO} \geq$ 1.6 nm (Fig. 1G). Note that the oscillation of the net magnetization was directly picked up from the magnetization value. This underscores the intrinsic nature of the oscillation in our superlattices, distinguishing it from the oscillations observed in the exchange bias reported by other studies (*31-34*). For example, SRO/NM-M/SRO heterostructure has shown exchange bias possibly originating from the RKKY interaction (*35*), yet *M* (*H*) curves of the SRO/STO superlattices do not show any exchange bias. Superlattices with $t_{SRO}$ = 0.8 nm or less show negligible in-plane *M* (*T*) possibly owing to the reduced dimensionality (*24*). But *M* (2 K) of superlattices with sub-atomic unit cells $t_{STO}$ control also presents a portion of the magnetic oscillation (Fig. S7). Fig. 1F further shows that the *M* (*H*) curve for the superlattices with thick STO layer (*i.e.*, $[6|18]_{10}$ superlattice, $t_{STO}$ = ~7 nm) returns to conventional single FM hysteresis loop with nearly the same saturation magnetization to the [6|6]$_{10}$ superlattice. The downturn of $M$ ($T$) with decreasing $T$ also disappears for the [6|18]$_{10}$ superlattice. These results indicate that an unconventional IEC plays a key role in the exotic AFM-like magnetic ground state of the superlattices with thin STO layers.

We fabricated more than 350 superlattices and measured the in-plane magnetization for more than 40 of those samples and achieved excellent reproducibility in both structural and magnetic properties (Fig. S8). This confirms that the $t_{STO}$-dependent oscillation of the net magnetization in the superlattices is intrinsic, and not originating from short-range ordering mechanisms such as spin-glass behavior. We note that the out-of-plane magnetization of the superlattices serves as the magnetic easy axis identical to the single SRO film (Figs. S4 and S5), but does not exhibit the peculiar oscillatory behavior. The oscillation is observed only for the relatively small, in-plane magnetization, demonstrating a stronger susceptibility to the unconventional IEC. The small in-plane magnetization further complicates the application of conventional IEC analyses using the $M$ ($H$) curves. Thus, to characterize the magnetic ground states of the superlattices, we focused on the in-plane magnetic behavior and spin structures at low-$H$ fields below 80 K.

**Synthetic spiral spin state**

The depth profile of the spin vector orientation along the in-plane direction of each FM layer supports the IEC-induced spiral spin ordering (Fig. 2), and resultant STO-thickness dependent oscillation of the net magnetization. Fig. 2A shows the PNR spectra of the [6|4]$_{10}$ superlattice measured at 5 K with 0.01 T in-plane $H$-field. R$^+$ (R$^-$) corresponds to the reflectivity measured with neutron spin parallel (antiparallel) to $H$. The superlattice peak at $Q = $ ~1.6 nm$^{-1}$ corresponds to the periodicity of the superlattice structures, consistent with the layer thicknesses obtained using X-ray reflectivity and scanning tunneling electron microscopy (Figs. S2, S3, and Table S1). Spin asymmetry (S. A. = (R$^+$ − R$^-$)/(R$^+$ + R$^-$)) exhibits a distinct structure at $Q = $ ~0.8 nm$^{-1}$ (half of the superlattice peak $Q$ value), indicating the existence of AFM ordering (Fig. 2B) (*19, 36, 37*). (We focused on the result of [6|4]$_{10}$ superlattice as the peak values at $Q = $ ~0.8 and ~1.6 nm$^{-1}$ provided the largest signal to noise ratio. We also did not carry out the spin-flip polarization because the PNR signals were estimated to be quite small (Fig. S9) (*36*).) To characterize the synthetic spin structure of the superlattice, we assumed that each FM SRO layer has a single spin orientation with an identical total magnetization, and the spin orientation of a neighboring layer has a relative angle difference of $\phi$, depending on $t_{STO}$ (Supplementary Text 1). The $\beta$- and $\gamma$- dependent $M$ (2 K) were fit to result in $\phi = $ ~160° between the FM layers for $t_{STO}$ = 1.6 nm, indicating a non-collinear spiral spin structure (Fig. S10). Fig. 2E visualizes the constructed spiral spin structure along the in-plane direction of the superlattice with nuclear (Fig. 2C) and magnetic (Fig. 2D) scattering length densities (SLDs). By examining numerous synthetic magnetic configurations, including the FM, sAFM, and spiral spin structures with varying $\phi$, we confirmed that the structure shown in Fig. 2E indeed leads to a consistent in-plane simulation result of the experimental PNR spectra (Figs. S12-S15, Table S2, and Supplementary Text 2). While the exact rotation of the spin directions from one SRO layer to the other can be a subject of debate based on our experimental and fitting errors, it is evident that collinear spin configurations including FM and sAFM cannot account for the observed experimental results.

Both the non-collinear spiral spin state and its $t_{STO}$-dependent behavior reflect the existence of a nonconventional IEC in the SRO/STO superlattice. Several mechanisms can be considered to understand the magnetic interaction, including the dipole-dipole interaction, magnetic anisotropy, RKKY, orange peel interaction, coupling to the defect states, interlayer Dzyaloshinskii-Moriya interaction, and chiral phonon induced IEC (*2, 18, 34, 37-41*). First, the $t_{STO}$-dependent oscillatory behavior can rule out the dipole-dipole interaction and magnetic anisotropy, as they would only impose monotonic $t_{STO}$-dependence (*36*). Second, the RKKY cannot explain the interaction across the NM-I STO spacers (Fig. S16 and Supplementary Text 3) (*18, 38*). Third, Néel suggested that a sizeable roughness at heterointerfaces can lead to an IEC across NM-I spacers (*38*), yet our superlattices have a roughness in the order of an atomic unit cell (~0.4 nm) (Fig. S3). Fourth, recent studies proposed that Schottky defects or metallic in-gap states can result in an IEC (*34, 37*), which is absent in our SRO/STO superlattices with suppressed charge transfer (See the Supplementary Text 4 for more detail) (*23, 24*). Fifth, interlayer Dzyaloshinskii-Moriya interaction can account for a non-collinear spiral spin state of synthetic magnetic layers. However, the Dzyaloshinskii-Moriya vector orientation should be defined along the in-plane direction of our experimental setup as was the case of other previously reported systems (*39, 40*), which is not consistent with the observed in-plane spiral spin structure.

**Chiral phonon mediated IEC**

The last candidate, *i.e.*, chiral phonon-spin coupling induced IEC is a plausible mechanism for elucidating the unconventional magnetic ground state, considering the facile formation and propagation of chiral phonons in the perovskite superlattices (*2, 13, 15, 41*). In particular, time-resolved optical spectroscopy has demonstrated that chiral phonons can dynamically mediate the long-range magnetic interaction across the NM-I spacer (*2*). They proposed a chiral phonon mediated-long-range spin exchange Hamiltonian, which explicates with the angular momentum transfer mechanism via chiral phonons and spin-orbit interaction, through the insulating layers. As the chiral nature of the phonon breaks the mirror symmetry, one can anticipate the existence of an anti-symmetric IEC.

For the SRO/STO superlattices, strong spin-phonon and spin-orbit coupling of SRO lead to the creation of the chiral phonons (*20-22*). In particular, the ferromagnetic molecular field of SRO layers breaks the degeneracy, enabling disparate population of the chiral phonons. The structural similarity of the perovskites and atomically sharp interfaces further facilitate the chiral phonon propagation, allowing chiral phonon-spin interaction in the neighboring SRO layers. Hence, we propose that spins in the FM layers can interact with each other through an NM-I spacer via chiral phonon-spin coupling, and the resultant unconventional magnetic ground state could be realized in the SRO/STO superlattices.

**Chiral phonon-spin coupling evidenced by phonon Zeeman effect**

The existence of chiral phonons and their coupling to spins in the SRO/STO superlattices was evidenced by $T$-dependent confocal Raman spectroscopy (see methods). Figure 3 shows a distinct split of a phonon mode at ~367 cm$^{-1}$ into two modes at ~358 (phonon $A$) and ~386 cm$^{-1}$ (phonon $B$) below $T_c$, in addition to the conventional phonon anomaly at $T_c$ originating from the strong phonon-spin coupling in SRO (*20, 21*). Lattice dynamical calculation indicated that the modes correspond to oxygen vibrations with orthogonal polarizations in the orthorhombic bulk SRO (*20*). A superposition of the two orthogonal linear phonons with phase difference can create chiral phonons (insets of Fig. 3B) (*15*), which are to be distinguished from the chiral phonons in hexagonal lattices (*3, 6*). The chiral phonon-spin coupling of SRO/STO superlattices was manifested by the phonon Zeeman splitting (*5*), which appeared without an external $H$-field. This reveals that the ferromagnetic molecular field in the SRO layer is strongly coupled to the chiral phonons (Fig. 3B). Whereas the two phonon modes are not degenerate in bulk SRO with orthorhombic structure, they should be degenerate in the thin film with a tetragonal structural symmetry, as shown for the high $T$ results. From the $T$-dependent behavior, we believe that the degeneracy lifting below $T_c$ is related to the opposite energy shift of the chiral phonons with opposite chirality under the ferromagnetic molecular field. We also note that SRO has no structural phase transition below $T_c$ (*21*). The relative splitting of phonon frequency ($\Delta\omega/\omega$) of SRO/STO superlattice is ~0.076, similar to that of 4$f$ rare-earth trihalides and Cd$_3$As$_2$ (*15, 42-45*). This indicates the strong effect of chiral phonon-spin coupling in SRO/STO superlattices. Although the population of phonon would decrease, the polarization of phonon can be enhanced with decreasing temperature, facilitating the magnetic interaction below $T_c$. In addition, the chiral phonon frequency ($\omega$) is closely associated with the spatial spin rotation period (~3.6 nm) of the SRO/STO superlattices. Assuming the phonon velocity of SRO ($v_s$) to be ~2-6 nm ps$^{-1}$ (*46*), the wavelength can be estimated as $2\pi v_s/\omega$ = ~1-3.6 nm for the chiral phonons showing the same order of magnitude to the periodicity of the synthetic magnetic oscillation. This correspondence suggests that chiral phonons created in the SRO layers can propagate across the NM-I STO spacers and induce IEC via chiral phonon-spin coupling.

**Discussion**

We have observed exotic spiral spin states with NM-I thickness-dependent oscillatory magnetic behavior, indicating the existence of IEC in atomically designed FM/NM-I heterostructures. The existence and the propagation of chiral phonons in the SRO/STO heterostructures would mediate the unconventional IEC via chiral phonon-spin coupling. We note that the microscopic mechanism of the unconventional IEC remains to be clarified and future theoretical studies are necessary. Nevertheless, the experimental observation itself provides a general intuition for understanding the emergent magnetic quantum phenomena, and the atomic-scale approach inspires the future combination of spintronics and phononics.

## Materials and Methods
### Superlattice growth
We chose the SRO and STO heterostructures since this system highly suppresses the charge transfer between SRO/STO interfaces, demonstrating intrinsic low-dimensional SRO layers (*23, 24*). We deliberately synthesized the $[\alpha|\beta]_\gamma$ superlattices, in which $\alpha$-atomic unit cells layers of SRO and $\beta$-atomic unit cell layers of STO are systematically repeated for $\gamma$ times along the growth direction, using pulsed laser epitaxy on (001) STO substrates. We controlled the number of atomic unit cells in the superlattices by employing a customized automatic laser pulse control system programmed using LabVIEW. The superlattice period ($d_{SL}$) was characterized by Bragg's law as $d_{SL} = \lambda/2 (\sin\theta_n - \sin\theta_{n-1})^{-1}$, where $\lambda$, $n$, and $\theta_n$ are the wavelength of the X-ray (0.154 nm for Cu K-$\alpha_1$), the order of superlattice peaks, and the $n$th-order superlattice peak position, respectively. We deduced the thickness of SRO and STO layer utilizing XRR simulations, as shown in Fig. S2. We ablated stoichiometric ceramic targets using a KrF laser (248 nm, IPEX868, Lightmachinery) with a repetition rate of 5 Hz and a laser fluence of 1.5 J cm$^{-2}$. Both SRO and STO layers were deposited at 750 ºC and 100 mTorr of oxygen partial pressure for the stoichiometric condition of both materials. Note that atomically thin STO layers well-preserve their insulating behavior (Fig. S16).

### Structural characterization
X-ray reflectivity and $\theta$-$2\theta$ measurements were carried out by using high-resolution X-ray diffraction (HRXRD) of Rigaku Smartlab and PANalytical X'Pert X-Ray Diffractometer. We estimated the thickness of the superlattice period using Bragg's law as, $\Lambda = \frac{\lambda}{2}(\sin\theta_n - \sin\theta_{n-1})^{-1}$, where $\Lambda$, $n$, $\lambda$, and $\theta_n$ are the period thickness, superlattice peak order, X-ray wavelength, and $n$th-order superlattice peak position, respectively. All the superlattices exhibit a small thickness deviation, below 1 atomic unit cell (~0.4 nm), corresponding to the atomic step-size of the substrate. Atomic-scale imaging of SRO/STO heterostructure was performed on a spherical aberration-corrected scanning tunneling electron microscopy (STEM, ARM200CF, JEOL) working at 200 kV with high-angle annular dark-field (HAADF) imaging mode. The incident electron probe angle was ~23 mrad that translates to a probe size of ~0.78 Å. The angle range of the HAADF detector was 70–175 mrad. Cross-sectional thin samples for STEM observation were prepared by a dual-beam focused ion beam system (FIB, FEI Helios Nano Lab 450) and the following low-energy Ar ion milling at 700 V (Fischione Model 1040, Nanomill) was conducted for 15 min to eliminate damaged surface layers from heavy Ga ion beam milling in the FIB system.

### Magnetization measurements
$T$- and $H$-dependence of magnetization behavior were measured using a Magnetic Property Measurement System (MPMS, Quantum Design). Field-cooled $M(T)$ curves were obtained from 300 to 2 K with 0.01 T of $H$-field. The $M(H)$ curves were obtained at 5, 50, and 85 K.

**Polarized Neutron Reflectivity**
PNR experiments were performed on the Magnetism Reflectometer at the Spallation Neutron Source at Oak Ridge National Laboratory (SNS, ORNL). The measurements were performed in closed-cycle refrigerator systems with an external $H$-field by using a Bruker electromagnet. We used highly polarized neutrons (polarization efficiency of 99 to 98.5 %) with wavelengths ($\lambda$) within a band of 2 to 8 Å. PNR spectra for two neutron polarizations ($R^+$ and $R^-$) were obtained by utilizing time-of-flight method, where a collimated polychromatic neutron beam impinges a sample at a grazing incidence angle ($\delta$). The reflectivity signal was recorded as a function of wave vector transfer, $Q = 4\pi\sin(\delta)/\lambda$ (*47*). Nuclear and spin structures of superlattices were simulated by using GenX (*48*). To characterize the nuclear structure of the sample, we measured NR spectra at 300 K (Fig. S12A). Saturation magnetization values were estimated to 0.4 $\mu_B$/Ru using PNR spectra at 85 K with 1 T (> coercive field at 85 K) of in-plane $H$-field (Fig. S12B), consistent with the result from MPMS. PNR is a depth-sensitive vector magnetometry method (*49*). In saturation, the layer magnetizations are fully magnetized parallel to the external field. In this case, only non-spin-flip reflectivity curves, $R^{++}$ and $R^{--}$ exist, which correspond to two orientations of neutron polarization. The reflectivity curves have well known characteristic features, like the oscillations determined by the total thickness and the Bragg peaks, corresponding to the periodicity of the superlattices. When the external magnetic field is released down to a small value of a guide field, the magnetizations in alternating SRO layers form a spiral structure, thus containing $M_x$ and $M_y$ components of magnetization. In this configuration the reflected signal will contain both non-spin-flip ($R^{++}$ and $R^{--}$, determined by the parallel component of the magnetization vector $M_y$) and spin-flip ($R^{+-}$ and $R^{-+}$, determined by the perpendicular component of the magnetization vector $M_x$) neutrons so that the resulting reflectivity curves will correspond to $R^+ = R^{++} + R^{+-}$ and $R^- = R^{--} + R^{-+}$. Our experiments were performed without spin-flip polarization analysis because the estimated experimental signal with spin-flip polarization is quietly small (Fig. S9). As it was demonstrated in ref. (*50*), without the full polarization analyses, the coupling angle between the magnetization vectors in the alternate magnetic layers can be effectively obtained through the data analyses by using the modulus of the magnetization vector obtained from the saturation and fitting only the alignment angles $\phi$ in the alternating SRO layers.

**Confocal Raman spectroscopy**
Raman spectra of SRO/STO superlattices were measured by utilizing a confocal micro-Raman (Horiba LabRam HR800) spectrometer using a HeNe laser with a wavelength of 632.8 nm (1.96 eV). To enhance the Raman cross-section of inelastic light scattering, we used the SRO/STO superlattice samples with 50 repetition number (*20*). We used a grating with 1800 grooves per mm and a focused beam spot size of ~5 μm. The power was maintained below ~0.3 mW to suppress any heating effects. With the backscattering geometry, we employed $z(xx)\bar{z}$ configuration to detect the $A$ and $B$ phonon modes, which are expected to exhibit a strong spin-phonon coupling. We note that an anomaly of phonon $B$ at $T_c$ represents the strong spin-phonon coupling in SRO/STO superlattices, consistent with previous SRO heterostructures (*20*, *21*). Lattice dynamical calculation suggested that the $A$ and $B$ phonon modes correspond to oxygen vibrations with orthogonal polarizations ($A_g$ and $B_{2g}$ modes, respectively) in the orthorhombic bulk SRO (*20*). To obtain high-quality Raman spectra, we precisely adjusted the *z*-directional beam-position to achieve the optimal focus on the superlattice samples (*50*).

**Junction transport measurement**

The junction transport measurement was performed on the superlattice with atomically thin STO layers. Tunneling transport geometry was employed using Nb-doped (0.5 wt %) STO substrate as the bottom electrode and Au (~500 μm in diameter) as the top electrode. The top electrode was patterned on the surface of superlattices utilizing RF sputtering with a shadow mask. *T*-dependent tunneling resistance of superlattices was recorded using a physical property measurement system (PPMS, Quantum Design) with an excitation current of 0.3 μA.

**Acknowledgments**

This research used resources at the Spallation Neutron Source, a DOE Office of Science User Facility operated by the Oak Ridge National Laboratory. We thank TSP. and JK for technical support of junction transports and KTK and SW for valuable discussions. We also thank Core Research Facilities, Pusan National University for MPMS.

**Funding:** Include all funding sources, including grant numbers, complete funding agency names, and recipient's initials. Each funding source should be listed in a separate paragraph such as:
National Research Foundation of Korea NRF-2021R1A2C2011340 (SGJ, WSC)
National Research Foundation of Korea NRF-2020K1A3A7A09077715 (SGJ, JK, SP, WSC)
National Research Foundation of Korea NRF-2020R1A2C1006207 (YMK)
National Science Foundation (DMR-1454200) (AS)
Alexander von Humboldt Foundation (Research Fellowship for Experienced Researchers) (AS)

**Author contributions:**
    Conceptualization: SGJ, WSC
    Methodology: SGJ, JK, AS, SP, HYJ, YMK, VL, WSC
    Data curation: SGJ, JK, AS, SP, YMK, VL
    Investigation: SGJ, SP, VL, WSC
    Validation: AS, SP, VL, TE, JHH, WSC
    Visualization: SGJ, WSC
    Supervision: WSC
    Project administration: WSC
    Funding acquisition.: WSC
    Writing—original draft: SGJ, WSC
    Writing—review & editing: SGJ, AS, SP, VL, JHH, WSC

**Competing interests:** : Authors declare no competing interests.

**Data and materials availability:** All data for samples and reference materials are available in the supplementary materials.

# Figures and Tables

## Table 1. The chiral phonon induced emergent spin states.

| Phenomena | Systems | References |
|---|---|---|
| Phononic helical edge state | Mechanical topological insulator | (*1*) |
| AC Stark effect | CdTe quantum well | (*2*) |
| Phonon Hall effect | Honeycomb lattice, cuprates, paramagneticdielectrics | (*10, 14, 16*) |
| Optically driven effective magnetic field | $ErFeO_3$ (Perovskite) | (*4*) |
| Topologically-induced viscosity split | Weyl Semimetals | (*5*) |
| Intervalley transfer of phonon angular momentum | $WSe_2$ (Honeycomb lattice) | (*6*) |
| Topological magnon-phonon coupling | Honeycomb lattices | (*7*) |
| Pseudogap phase of superconductor | Cuprates | (*8*) |
| Einstein–de Haas Effect | Honeycomb, Kagome, triangle, and square lattices | (*3, 11, 12*) |
| Dynamic multiferroicity | Perovskites | |
| Resonant magnon excitation | $ErFeO_3$ (Perovskite) | |
| Dzyaloshinskii-Moriya-type electromagnon | $TbMnO_3$ (Perovskite) | (*13, 15*) |
| Inverse Faraday effect | $DyFeO_3$ (Perovskite) | |
| Phonon Zeeman effect | Binary compounds, perovskites, transition metal dichalcogenides | |
| Entanglement of single-photon and phonon | $WSe_2$ (Honeycomb lattice) | (*9*) |
| Static interlayer exchange coupling | $SrRuO_3/SrTiO_3$ superlattice | Current study |

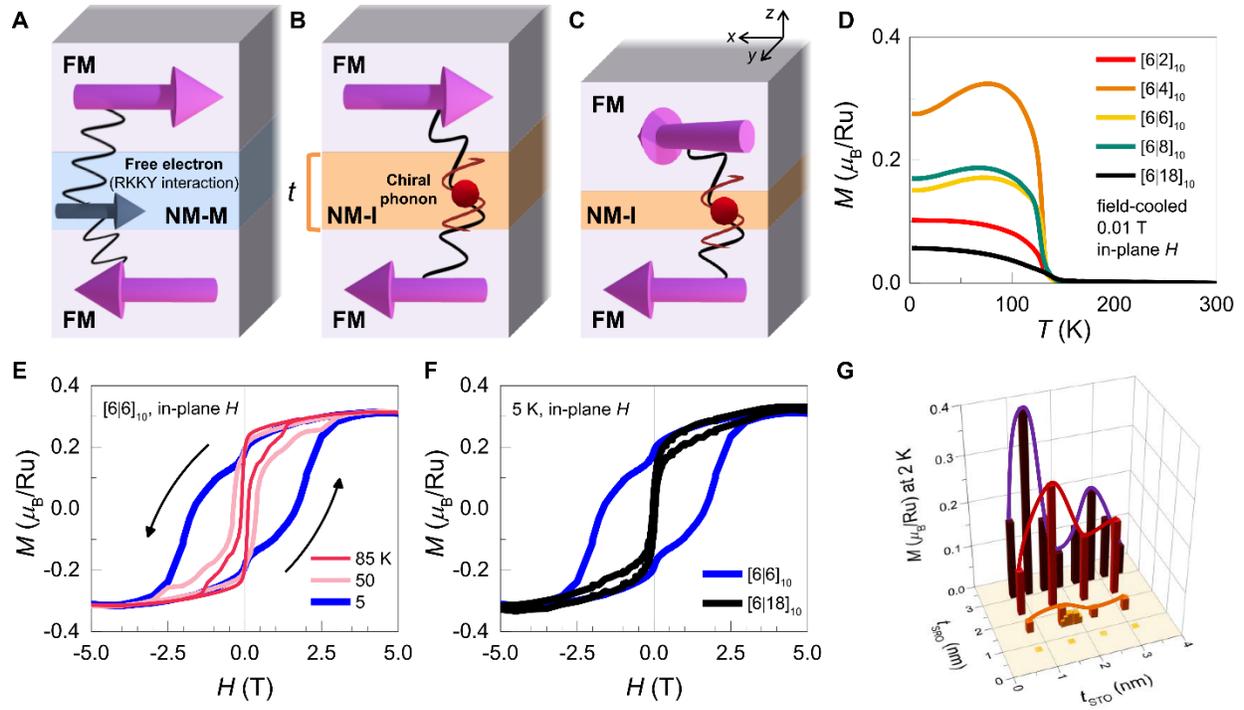

**Fig. 1. Chiral phonon-mediated IEC and resultant $t_{STO}$-dependent magnetic behavior of SRO/STO superlattices.** Schematic diagrams of IEC across (A) an NM-M spacer via free-electron mediated RKKY interaction and (B and C) NM-I spacers via chiral phonon-mediated interaction with different $t$. (D) In-plane $M$ ($T$) curves of SRO/STO superlattices with systematically changing $t_{STO}$. $T^*$ is marked with an arrow. (E) $M$ ($H$) curves of the superlattices at 85, 50, and 5 K. The arrows indicate the directions of the $H$-field. (F) $M$ ($H$) curves of the superlattices with different $t_{STO}$. The curves are measured at 5 K. (G) Oscillatory magnetic behavior of the $M$ value as a function of $t_{STO}$. The values were obtained at 2 K with 0.01 T of in-plane $H$-field. The solid lines are the guide to the eye.

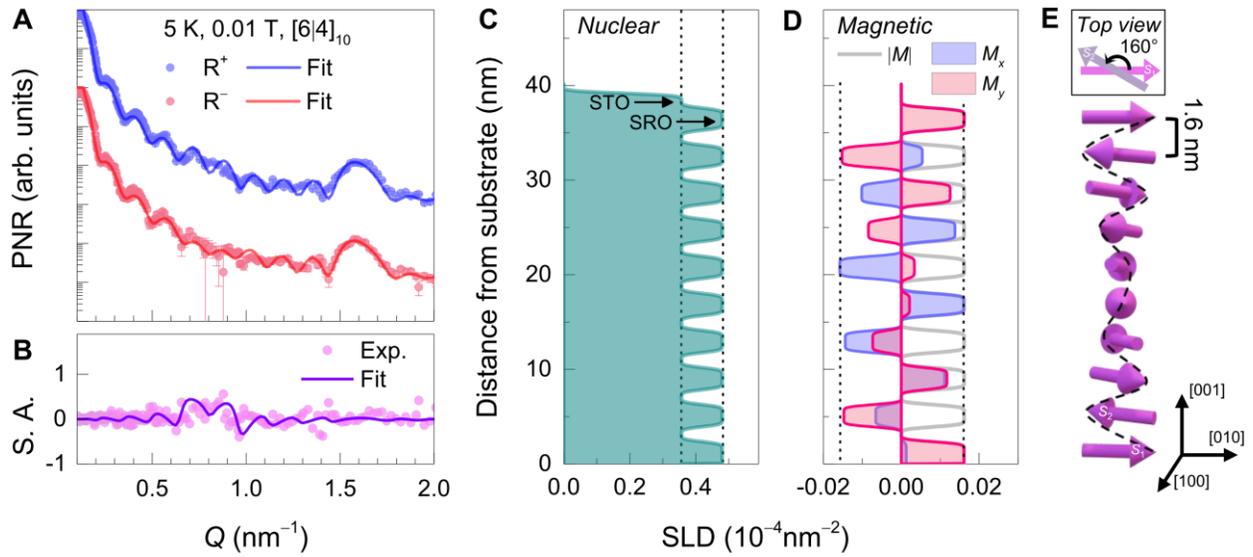

**Fig. 2. Non-collinear spiral spin state of the SRO/STO superlattice.** (A) PNR spectra for the spin-up ($R^+$) and spin-down ($R^-$) polarized neutrons and (B) S. A. for [6|4] superlattice at 5 K. The measurement was performed with 0.01 T of in-plane $H$-field. The symbols and solid lines indicate experimental data and fit using the model in (C-E), respectively. The error bars represent one standard deviation. (C) Nuclear SLD depth profile of the superlattice. (D) Magnetic SLD depth profile with $x$- and $y$-directional $M$ values ($M_x$ and $M_y$). Gray solid line is the absolute value of the total magnetic SLD for each SRO layer within the superlattice (~0.4 $\mu_B$/Ru). The vertical dashed lines in (C) and (D) are guides to the eye. (E) Schematic diagram of the spin configuration in the SRO/STO superlattice. The top view displays the $\phi = 160°$ between the magnetization directions ($S_1$ and $S_2$) in the two neighboring SRO layers when $t_{SRO} = 1.6$ nm. External $H$-field has been applied along the [010] direction ($y$).

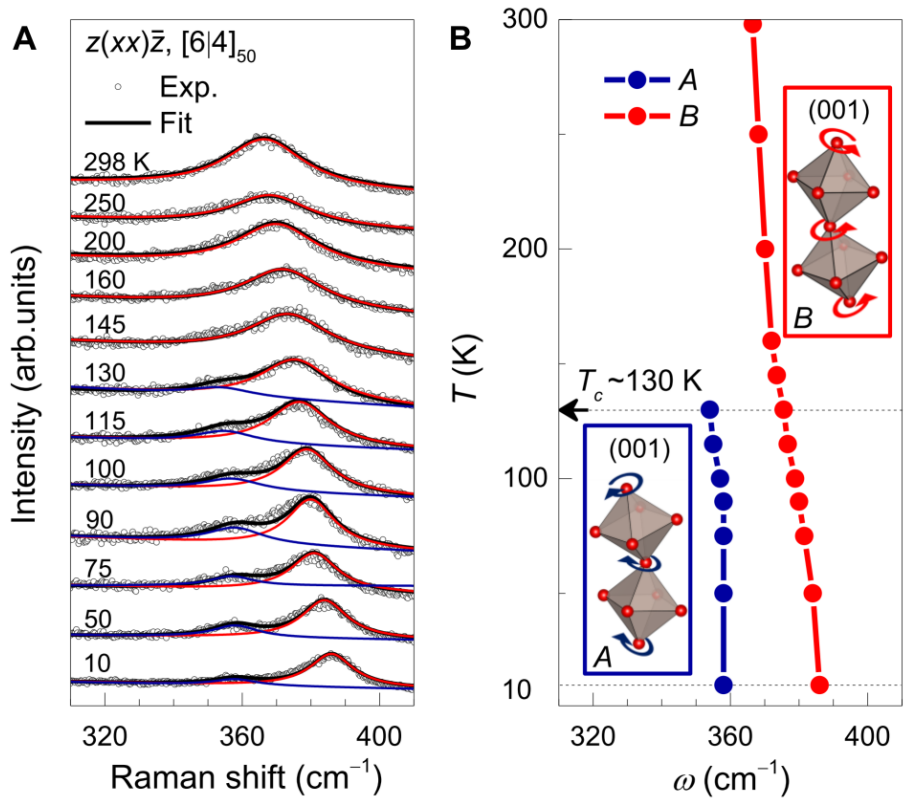

**Fig. 3. Chiral symmetry breaking in SRO/STO superlattices.** (A) $T$-dependent confocal Raman spectra in $z(xx)\bar{z}$ polarization of a [6|4] superlattice (The results of a [6|8] superlattices are in Fig. S17). The symbols and solid lines represent experimental data and fit for the Raman spectra, respectively. Blue and red solid lines correspond to the $A$ and $B$ phonons, respectively, as described in the insets of (B). (B) $T$-dependent $\omega$ splitting below $T_c$. The insets schematically represent the expected chiral phonon modes of SRO. The dotted horizontal lines indicate (up) $T_c$ of SRO and (down) the lowest $T$ (10 K) of the measurement.

**Supplementary Materials**

Supplementary Text 1 to 4
Figs. S1 to S17
Tables S1 to S2
References (*52* to *54*)

# Supplementary Materials for

## Unconventional interlayer exchange coupling via chiral phonons in synthetic magnetic oxide heterostructures

Seung Gyo Jeong, Jiwoong Kim, Ambrose Seo, Sungkyun Park, Hu Young Jeong, Young-Min Kim, Valeria Lauter, Takeshi Egami, Jung Hoon Han, and Woo Seok Choi1*.

*Corresponding author. Email: choiws@skku.edu

**This PDF file includes:**

Supplementary Text 1 to 4
Figs. S1 to S17
Tables S1 to S2
References (*52* to *54*)

**Supplementary Text**

Supplementary Text1. Characterization of the spiral spin structure of the SRO/STO superlattices using a phenomenological model.

To figure out the orientation of the spin vectors in each SRO layer within the superlattices and describe the oscillatory magnetic behavior of $M$ (2 K) values, we modeled synthetic spiral spin structures with different $\gamma$ (in [SRO|STO]$_\gamma$, $\gamma = 2$ (bilayer SRO), 3 (trilayer SRO), 4, and 5) and $t_{STO}$ (Figs. S8 and S10). Assuming that each SRO layer has a magnetization vector $M_i$, we estimate that neighboring SRO layers have the same amplitude but a relative angle difference ($\phi$) linearly dependent on $t_{STO}$. In order to calculate the total magnetization of the superlattice, we estimated the sum of projections of $M_i$ ($\phi$) along the $H$-field for each SRO layer, mimicking the result from MPMS, which only shows an average scalar magnetization value of the whole superlattice along the $H$-field direction. Assuming that the bottommost spin vector is directed along the $H$-field, $M_i$ ($\phi$) was defined as,

$$\sum_1^\gamma M_i(\phi) = \sum_1^\gamma [M_0 + M_1 \cos((\gamma - 1)\phi)], \quad (1)$$

where $M_0$ and $M_1$ are independent of $\phi$. Fig. S10b exhibits a $\phi$-dependent sum of squared error (SSE) values, calculated by [(Simulated $M$) – (Measured $M$)]$^2$. The spiral spin structure model with $\phi = \sim160$ and $\sim200°$ have the lowest SSE values for $t_{STO}$ = 1.6 nm (Inset of Fig. S10B). $M_0$ and $M_1$ values are 0.070 and 0.629 $\mu_B$/Ru, respectively. While spin models with both $\phi$ values well reproduce the $\gamma$-dependence of $M$ (2 K) (Fig. S10C), the oscillatory magnetic behavior of $M$ (2 K) as a function of $t_{STO}$ can be better explained by the spiral spin structure with $\phi = \sim160°$ (Fig. S10D). Thus, we conclude that the spiral spin structure with $\phi = \sim160°$ is the most reasonable spin configuration to describe the magnetization results obtained by MPMS. We emphasize that the proposed magnetic configurations do not yield a zero remnant magnetization. See Fig. S11 for clear visualization of the spin configurations. It should also be noted that this spiral spin structure coincides with the PNR measurements as shown below.

Supplementary Text 2. Analysis of PNR data.

We first measured the NR spectra at 300 K without $H$-field to examine the atomic structure of the superlattice (Fig. S12A). The result is highly consistent with the XRR result (Fig. S2 and Table S1). We then estimated the saturation magnetization values of the superlattice by measuring the PNR spectra at 85 K with 1 T (> $H_c$ at 85 K) of in-plane $H$-field (Fig. S12B). The experiment was done without spin-flip polarization analysis, $R^+ = R^{++} + R^{+-}$ and $R^- = R^{--} + R^{-+}$, where non-spin-flip $R^{++}$ ($R^{--}$) is proportional to $M_y = M \sin(\phi)$ and $R^{+-}$ ($R^{-+}$) is proportional to $M_x = M \cos(\phi)$ (*34, 41, 46-48, 53*). The estimated saturation magnetization value was $\sim0.4$ $\mu_B$/Ru, consistent with the MPMS. Assuming that the scalar magnetization value of each SRO layer is $\sim0.4$ $\mu_B$/Ru, we carefully fit the PNR spectra at 5 K with a 0.01 T of in-plane $H$-field utilizing three different spin structures, *i.e.*, a collinear FM, a collinear sAFM, and a non-collinear spiral spin model with $\phi = 160°$ as shown in Supplementary Text 1. The simulated PNR spectra using collinear spin models have clear distinctions from our PNR data: (1) The strong dip at $Q = \sim0.8$ nm$^{-1}$ (half of the superlattice peak $Q$ value) in $R^-$ of the collinear sAFM model simulation (Fig. S13B) and (2) the large differences at $Q = \sim1.6$ nm$^{-1}$ (the superlattice peak $Q$ value) between $R^+$ and $R^-$ of the collinear FM model simulation (Fig. S13C) cannot account for the experimental PNR spectra. On the other hand, the non-collinear spiral spin structure with $\phi = \sim160 \pm 5°$ well-describes our PNR data (Fig. S14), consistent with the analyses from the magnetization results of MPMS as described in Supplementary Text 1 and Fig. S10. In addition, we tested the FM contribution to our PNR spectra by combining the two different spin models, *i.e.*, FM ($M_0$) and spiral spin structures with $\phi = \sim 160°$ ($M_1$) (see Supplementary Text 1, Fig. S15). Table S2 shows that SSE of three different spin models with different ratios between $M_0$ and $M_1$ results in nearly the same SSE values. While this result implies that some FM contribution might exist in our superlattice system at 5 K, the existence of non-collinear spiral spin structure in the SRO/STO superlattices is undeniable, consistently supported by both the MPMS and PNR measurements and analyses.

Supplementary Text 3. Estimation of the RKKY-induced magnetic oscillation.
We estimated the RKKY interaction induced oscillation wavelength ($\lambda_{RKKY}$), assuming that the STO spacer layer has a finite number of itinerant carriers possibly from unintentionally introduced oxygen vacancies. We note that such speculation is unlikely, as the junction current across the superlattice shows an insulating *T*-dependent behavior as shown in Fig. S16. Junction transport of [6|2]$_{10}$ was performed using Nb-doped (0.5 wt %) STO substrate as the bottom electrode and Au (~500 μm in diameter) as the top electrode, as already mentioned in the method section in our manuscript. A few kΩ of tunneling resistance was observed for the ultrathin STO layer thickness (~8 nm) within the [6|2]$_{10}$ superlattice. If we convert the resistance into resistivity, the value becomes in the order of ~$10^6$ Ω cm, which is compared to that of wood. This is a large enough resistivity considering quantum tunneling is possible in the superlattice. Note that we used high oxygen partial pressure of 100 mTorr to grow the superlattices. Numerous experiments from other groups consistently confirm that 100 mTorr of oxygen partial pressure during growth results in the most stoichiometric and insulating STO thin films. The insulating behavior of the STO layer within our superlattices was further confirmed from optical, XAS, DFT, and STEM-EELS-EDX analyses performed in one of our previous studies on the superlattices (*24, 25*). Particularly, the Ti $L_3$-edge XAS spectra revealed the prevalence of only the Ti$^{4+}$ valence state, indicating no (unintended) itinerant carriers in the STO layers. We further note that the Ti$^{4+}$ valence state of the superlattice is independent of the STO thickness. Although such speculation is highly unlikely, by considering the carrier density of the STO spacer layer ($n_{STO} = \sim 10^{18}$ cm$^{-3}$) (*54*), the $\lambda_{RKKY}$ was determined by,

$$\lambda_{RKKY} = \frac{\pi}{k_F} = \frac{\pi}{(3\pi n_{STO})^{\frac{1}{3}}}. \tag{2}$$

The estimated $\lambda_{RKKY}$ was ~14.9 nm, which is an order of magnitude larger than the observed magnetic oscillation periods (~2 nm) in the SRO/STO superlattices. Thus, the RKKY interaction cannot account for the magnetic behavior through the NM-I STO spacer.

Supplementary Text 4. Extended discussion of possible IEC through a NM-I layer.
(1) Schottky defects-induced IEC
It is noteworthy that there are a few recent experimental reports on magnetic oscillations across insulating spacers (*31-34*). However, interpretations of the magnetic oscillation are often controversial, and most of them conceive unintentional charge carriers remaining in the NM-I spacer layer. More importantly, previous observations rely on the change in the macroscopic exchange bias, stressing the role of extrinsic magnetic pinning layer. In particular, a recent report speculated that Schottky defects in NM-I spacer within a polar heterostructure can induce an oscillation of exchange bias via defect-induced electron hopping, by showing a defect-induced Raman active mode for the NM-I spacer, with strong magnetic field dependence (*34*). Note that the reference does not provide any rigorous origin of the behavior, and it does not indicate any phonons playing a role. Our SRO/STO superlattices are clearly different. SRO/STO superlattices exhibit highly suppressed charge transfer at the interfaces, evidenced by various experiments and theoretical calculations (*23-25*). There is no exchange bias effect in the $M$ ($H$) curves of the SRO/STO superlattices, indicating the absence of extrinsic magnetic pinning layers (Numerous experiments consistently showed no exchange bias in SRO/STO heterostructures (*21, 24-27*).). Furthermore, observed phonon modes in the SRO/STO superlattices are octahedral distortion-related intrinsic phonon modes in the SRO layer, and not defect-induced modes of the NM-I spacer. Finally, our system is nominally non-polar, and hence, cannot be understood in terms of Schottky defect-induced IEC (*34*). The identical $A$-site (Sr) ions in our SRO/STO superlattices lead to a nominally symmetric interface.

(2) Interfacial magnetic coupling
The absence of exchange bias in $M$ ($H$) curve of the SRO/STO superlattice indicates that there is no extrinsic pinning layer in the superlattices, especially near the interfaces. Moreover, the $t_{STO}$-dependent oscillation of magnetization cannot be understood from any interfacial magnetic coupling. $M$ ($H$) curve for the superlattices with thick STO layer (*i.e.*, [6|18]$_{10}$ superlattice, $t_{STO}$ = ~7 nm) returns to conventional single FM hysteresis loop with almost the same saturation magnetization of [6|6]$_{10}$ superlattice. It strongly supports that the interfacial magnetic coupling cannot account for the unconventional IEC in SRO/STO superlattices.

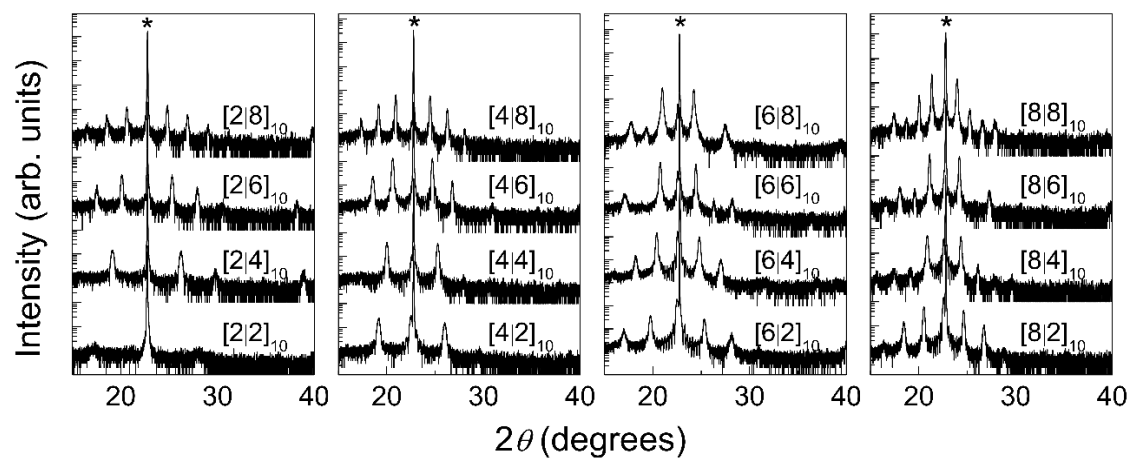

**Fig. S1. XRD $\theta$-$2\theta$ scans of $[\alpha|\beta]_{10}$ superlattices with different $\alpha$ and $\beta$.** Clear superlattice Bragg peaks show the atomically well-defined periodicities of superlattices. The asterisk (*) indicates the STO (001) substrate peak.

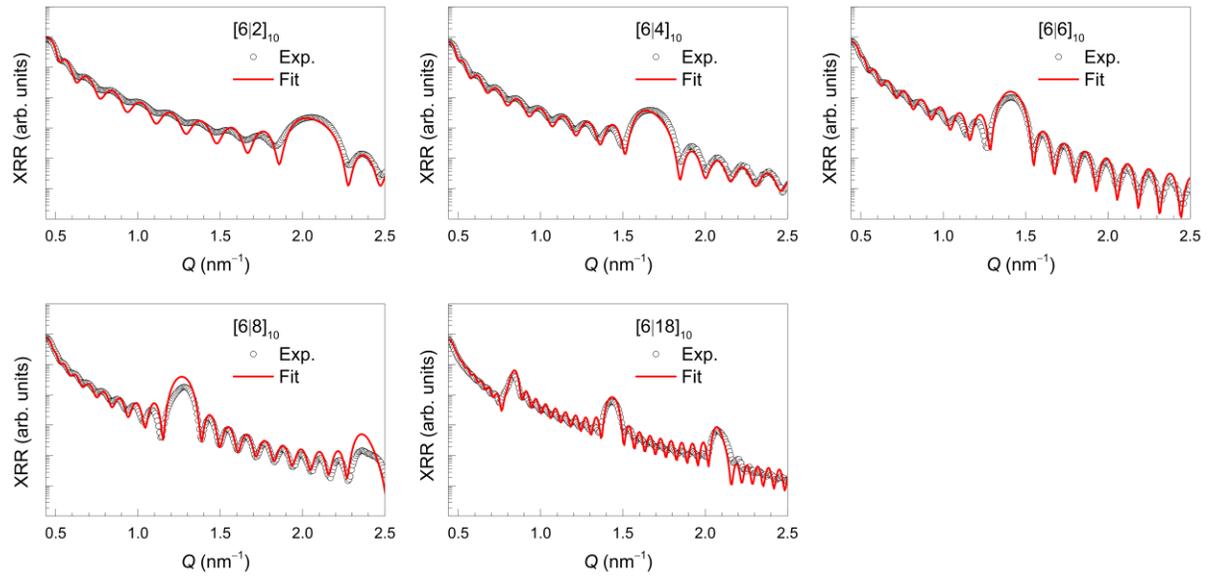

**Fig. S2. X-ray reflectivity (XRR) of [6|*β*]₁₀ superlattices with different *β*.** The XRR result shows atomically controlled periodicities of the superlattices with a small deviation of the thickness < 0.1 nm. Thus, we believe magnetic inhomogeneity would be small, if any, and further, it would not disrupt the NM-I spacer thickness-dependent unconventional magnetic behavior. The symbols and solid lines are the experimental data and fit of the XRR data.

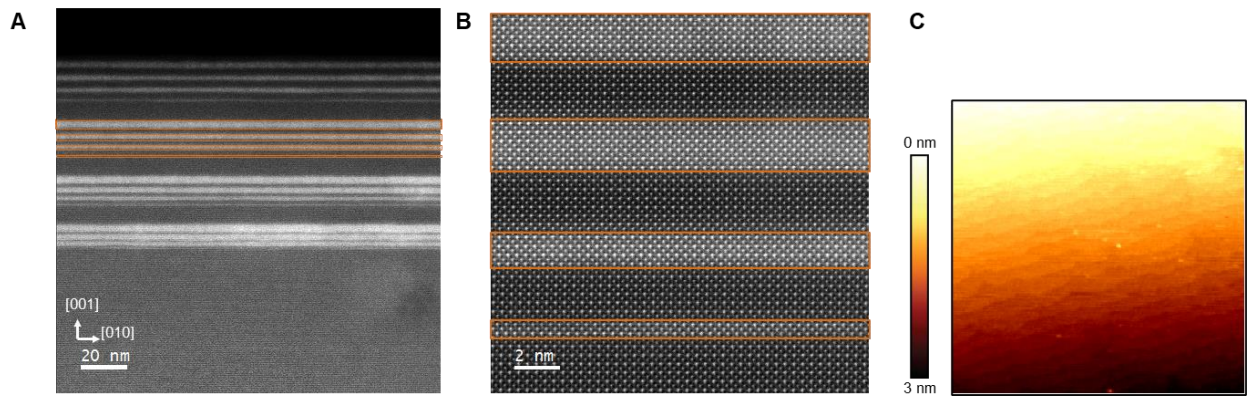

**Fig. S3. Atomically sharp interface and surface of SRO/STO heterostructures (not a superlattice).** Cross-sectional HAADF-STEM images of the SRO/STO heterostructures (not a superlattice) in (A) low and (B) high magnifications. The number of atomic layers was clearly visualized to confirm the designed structure, and the interfaces were proven to be atomically sharp. The bright layers enclosed by orange rectangles indicate the SRO layers. The STEM images manifest a coherent superlattice with a fully-strained state (*24, 25*). (C) Surface topography of SRO/STO heterostructures with an atomically flat step-and-terrace structure recorded using atomic force microscopy. This image was obtained in $5 \times 5$ μm$^2$ scales using contact mode.

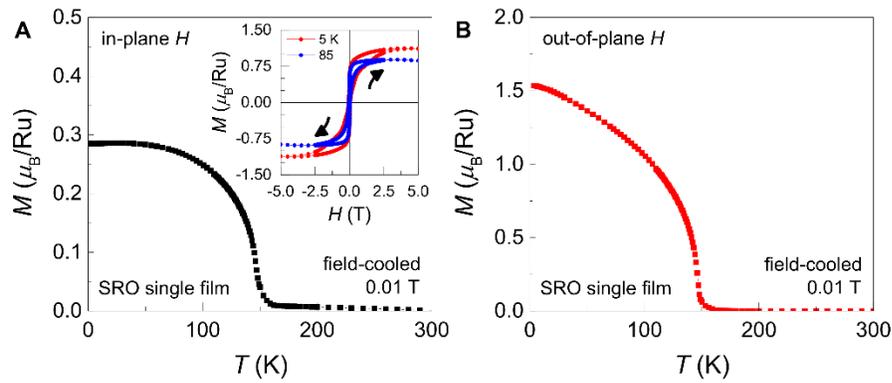

**Fig. S4. *M* (*H*) and *M* (*T*) curves of an SRO single film.** (A) In-plane and (B) out-of-plane *M* (*T*) curves of an SRO single film measured with 0.01 T of *H*-field using a field-cooled method. The inset shows *H*-dependent in-plane magnetization of an SRO single film. The arrows indicate the direction of the *H*-field. This result shows the nearly second-order FM transition of SRO, consistent with previous studies (*52*), and that the magnetic easy axis of SRO single films lies along the out-of-plane direction. We also note that the magnetic anisotropy field of SRO is in the order of 10 T, indicating out-of-plane magnetic moment cannot rotate along the in-plane direction under a few T of the in-plane magnetic field applied.

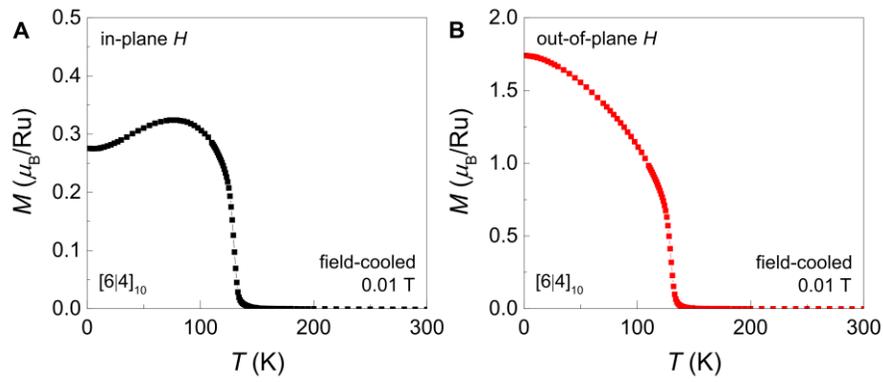

**Fig. S5. Anisotropic *M* (*T*) curves of an SRO/STO superlattice.** (A) In-plane and (B) out-of-plane *M* (*T*) curves of [6|4]$_{10}$ superlattice measured with 0.01 T of *H*-field using a field-cooled method. We note that the magnetic easy-axis of the superlattice is still out-of-plane, but the $t_{STO}$-dependent magnetic oscillation was observed only along the in-plane direction.

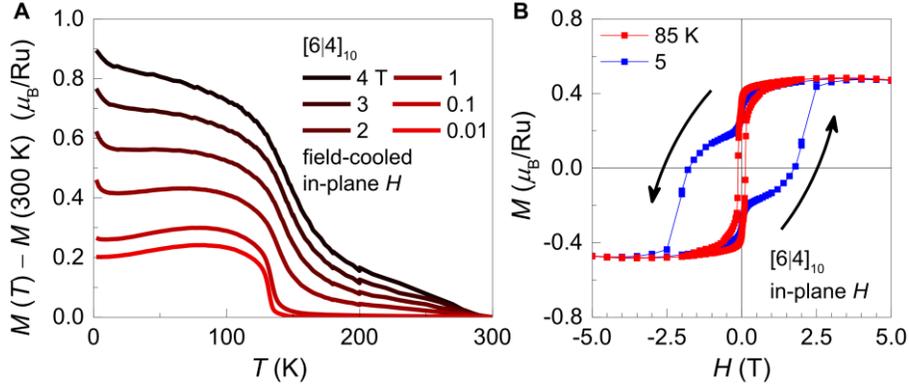

**Fig. S6. *M* (*T*) curves of an SRO/STO superlattice at different *H*-field.** (A) *M* (*T*) curves of a [6|4]$_{10}$ superlattice recorded at various in-plane *H*-field. The field-cooled method was used. Because the diamagnetic of STO with a high magnetic field show a large negative offset of the *M* value, we displayed *M* (*T*) – *M* (300 K) value. (B) In-plane *M* (*H*) curves measured at 5 and 85 K. The arrows indicate the direction of *H*-field.

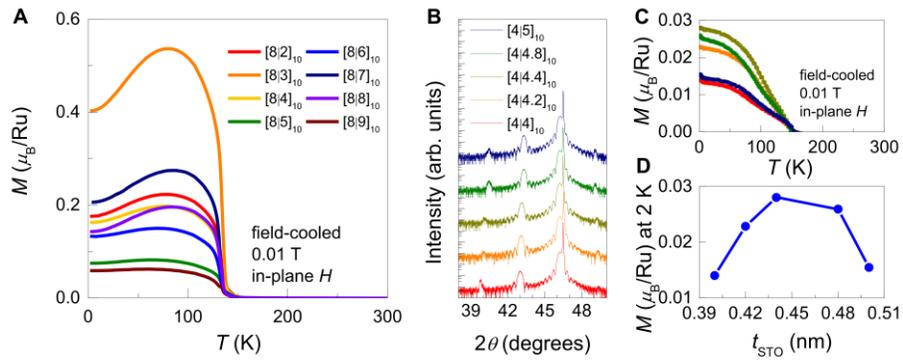

**Fig. S7. STO-thickness dependent oscillations in [8|β]₁₀ and [4|β]₁₀ superlattices.** (A) *M* (*T*) curves of the [8|β]₁₀ superlattices. (B-D) Results from sub-u.c. controlled [4|β]₁₀ superlattices. (B) XRD *θ*-2*θ* scans and (C) *M* (*T*) curves of sub-u.c. controlled [4|β]₁₀ superlattices. (D) *M* (2 K) values extracted from (C). These results support the oscillatory magnetization behavior in the SRO/STO superlattices. Note that this sample set with 4 u.c. SRO thickness shows lower absolute magnetization values compared to other superlattices shown in the main text, possibly originating from the thickness deviation of the SRO layer.

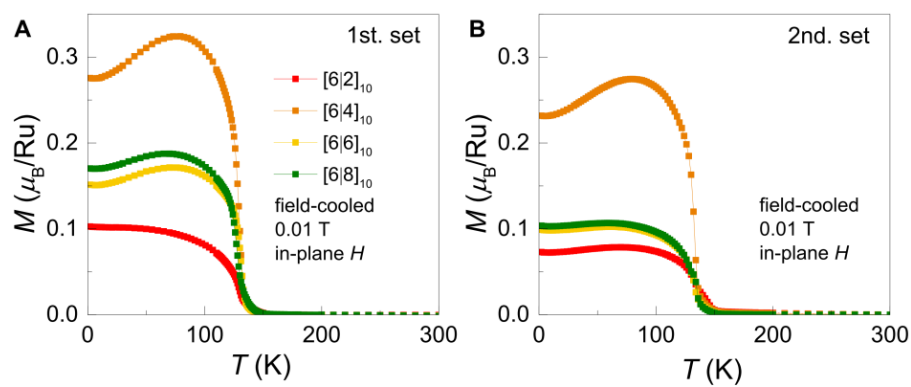

**Fig. S8. Reproducibility of the magnetic behavior of the superlattices.** (A) and (B) Two different sample sets of the superlattices with the same configuration. Both consistently show the $t_{STO}$-dependence, indicating good reproducibility of the oscillatory result. $M(T)$ curves were measured with 0.01 T of in-plane $H$-field using the field-cooled method.

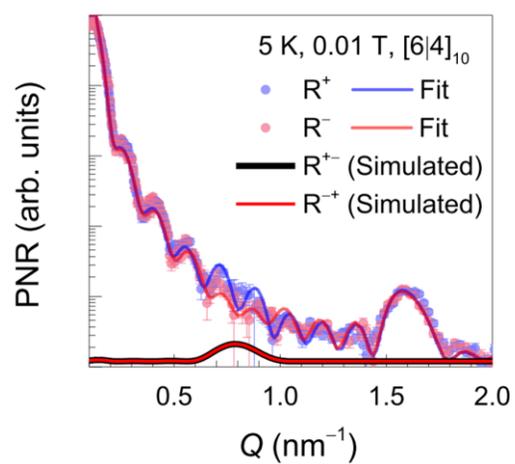

**Fig. S9. Estimated PNR spectra with spin-flip polarizations.** Estimated PNR spectra with spin-flip polarizations ($R^{+-}$ and $R^{-+}$) show the small PNR signal.

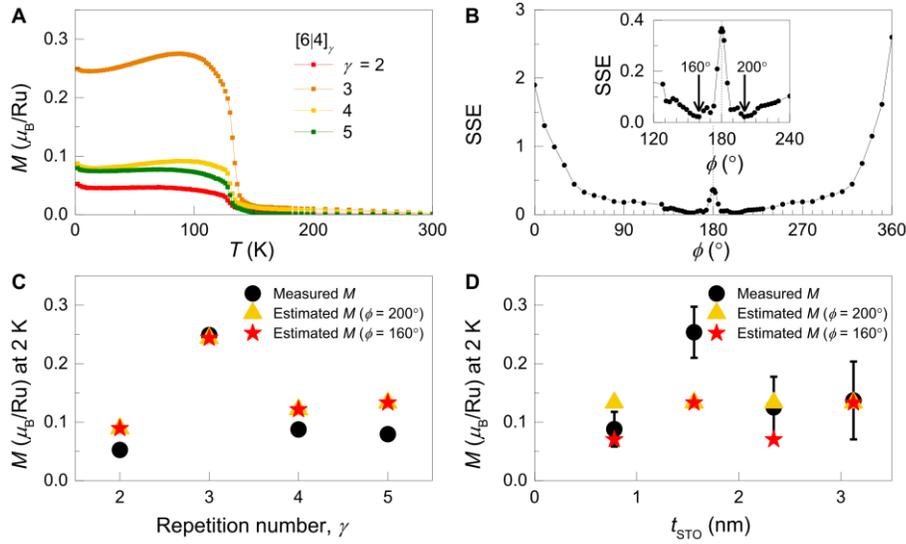

**Fig. S10. Characterization of the spin vectors for the SRO/STO superlattices.** (A) $M(T)$ curves of $[6|4]_\gamma$ superlattices with different $\gamma$. $M(T)$ curves were measured with 0.01 T of in-plane $H$-field using the field-cooled method. (B) $\phi$-dependent SSE values (see Supplementary Text 1). The dotted vertical line indicates $\phi = 180°$, indicating the sAFM configuration. The inset shows a magnified region near $\phi = 180°$. The arrows indicate the lowest SSE values at $\phi = 160$ and 200°. (C) $\gamma$- and (D) $t_{STO}$-dependent $M$ (2 K) values and the estimated magnetization values using the model described in Supplementary Text 1. The error bars indicate the deviation from the measurements.

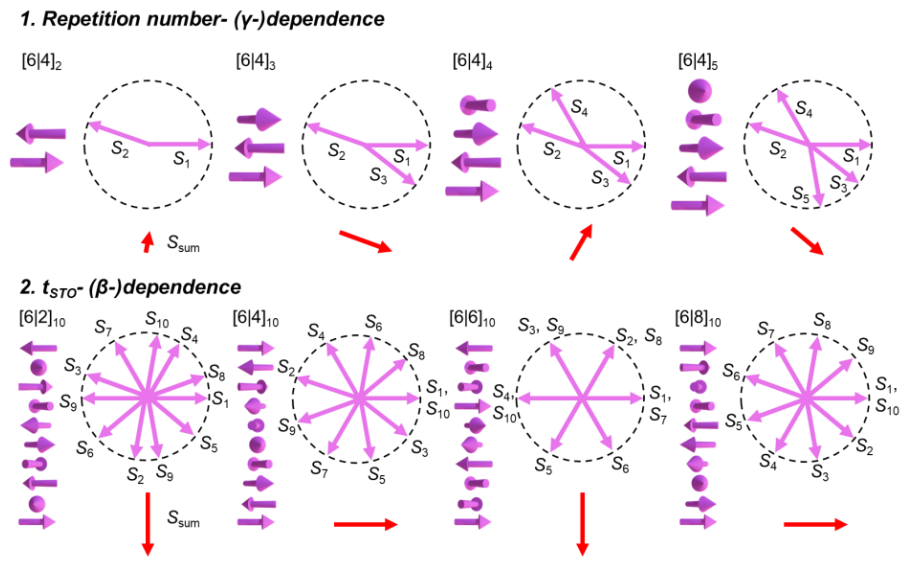

**Fig. S11. Schematic representation of the in-plane spin vector configurations and net magnetization in the SRO/STO superlattices.** $S_i$ represents the in-plane spin of the SRO layer starting from the bottommost layer ($S_1$). Assuming that each SRO layer has an in-plane magnetization vector $M_i$, we estimate that neighboring SRO layers have the same amplitude, but different orientation (see Supplementary Text 1). The red arrows denote the summation of vectors, $S_{sum}$, indicating the non-zero net in-plane magnetization that oscillates with $t_{STO}$. (1) Repetition number- ($\gamma$-) and (2) $t_{STO}$- ($\beta$-) dependent $S_{sum}$ are shown for comparison.

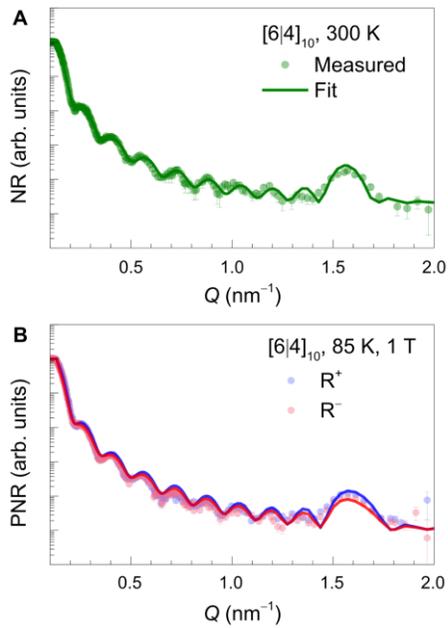

**Fig. S12. PNR spectra were measured at different *T*.** (A) Unpolarized neutron reflectivity (NR) spectra at 300 K. (B) PNR spectra at 85 K with an in-plane *H*-field of 1 T, where $R^+$ and $R^-$ indicate the PNR spectra for spin-up and spin-down neutrons, respectively. We effectively reduced the number of fitting parameters in our PNR fitting. Initially, the structural fitting parameters were fixed based on our XRR and NR measurements. Then, we estimated the saturation magnetization values of the superlattice by measuring the PNR spectra at 85 K with 1 T ($> H_c$ at 85 K) of in-plane *H*-field.

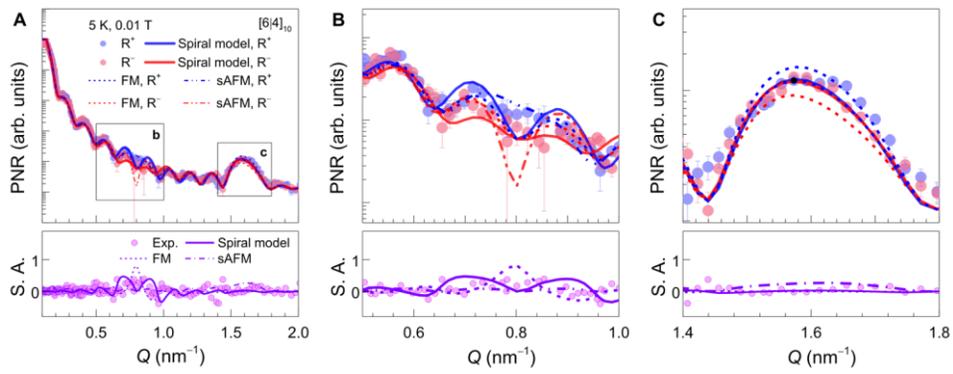

**Fig. S13. Analysis of the PNR spectra.** (A) PNR spectra with S. A. results for three models corresponding to three different scenarios for spin configuration (see Supplementary Text 2). Magnified PNR spectra for (B) $Q = 0.5–1.0$ nm$^{-1}$ and (C), $Q = 1.4–1.8$ nm$^{-1}$ region.

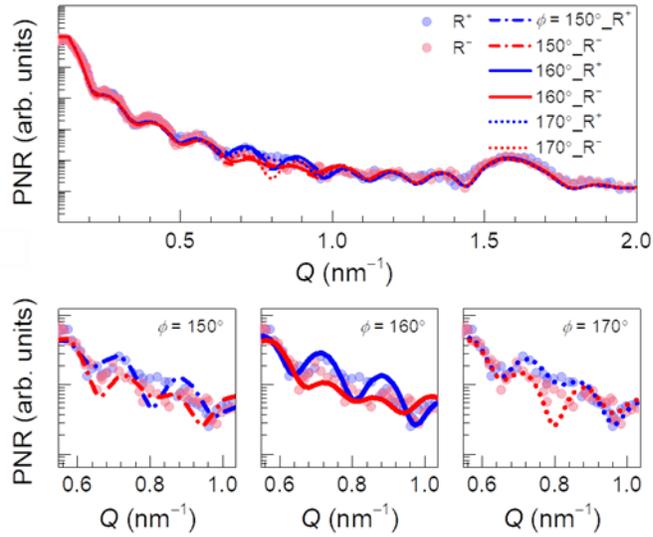

**Fig. S14. Simulation of non-collinear spiral spin structure with varying $\phi$.** To confirm the sensitivity of $\phi$ in the spiral spin structures of the superlattices, we performed the simulation with three different values of $\phi$ ($\phi = 150°$, $160°$, and $170°$). The bottom panels show the enlarged portion near $Q = 0.8$ nm$^{-1}$. The result from spin model with $\phi = 150°$ shows a large discrepancy to the experimental spectra for the $R^+$ polarization (blue), whereas the result from spin model with $\phi = 170°$ shows a large discrepancy to the experimental spectra for the $R^-$ polarization (red). The result from spin model with $\phi = \sim 160°$ best describes our experimental PNR data, consistent with the magnetization analyses based on the results of MPMS.

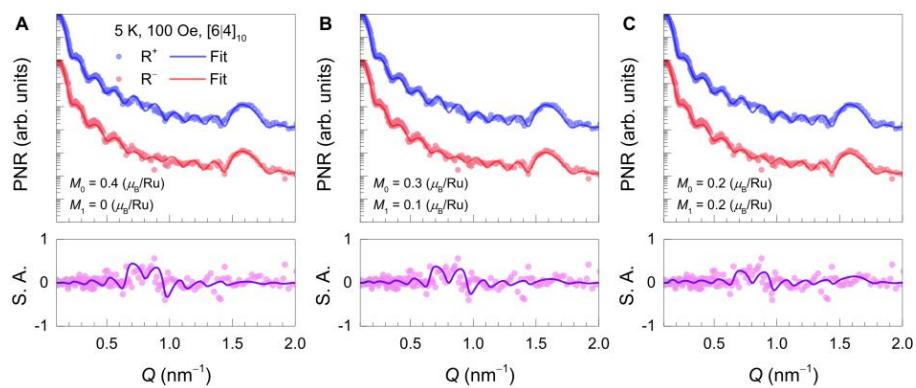

**Fig. S15. Partial FM contribution in the PNR spectra.** (A-C) PNR spectra and S. A. results for three different spin models, including different contributions of the FM component, $M_0$ (see Supplementary Text 2).

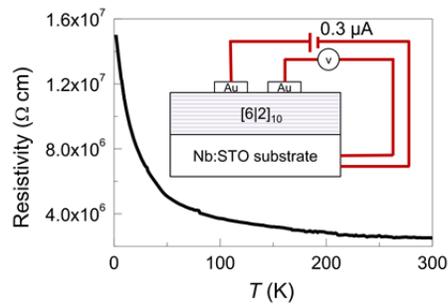

**Fig. S16. Junction transport property of the superlattice with atomically thin STO layer.** *T*-dependent resistance measured using junction transport geometry of a [6|2]$_{10}$ superlattice. 0.3 μA of the excitation current was used. The superlattice even with the thin STO layers (~0.8 nm) clearly shows an overall insulating *T*-dependence. Since SRO layer is metallic, the STO layer should be insulating to show the overall insulating *T*-dependence. The inset schematically shows the measurement configurations. See Supplementary Text 3 for more discussion.

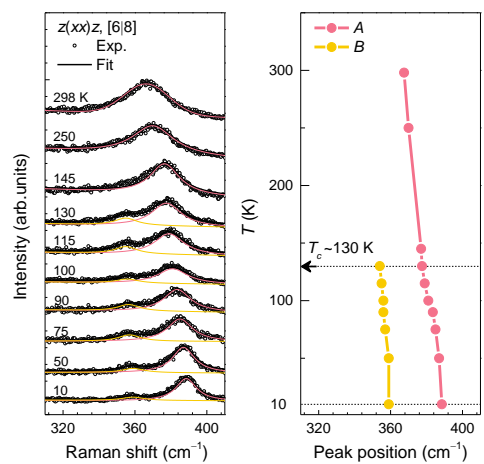

**Fig. S17. *T*-dependent confocal Raman spectra in $z(xx)\bar{z}$ polarization of a [6|8] superlattice.** The symbols and solid lines represent experimental data and fit for the Raman spectra, respectively. Pink and yellow solid lines correspond to the *A* and *B* phonons. *T*-dependent $\omega$ splitting below $T_c$. The dotted horizontal lines indicate (up) $T_c$ of SRO and (down) the lowest *T* (10 K) of the measurement.

| Samples | Designed thickness (nm) | Measured thickness (nm) | | |
|---|---|---|---|---|
| | | XRR | PNR | STEM |
| $[6\|2]_{10}$ | 31.366 | 31.256 | - | 32.590 |
| $[6\|4]_{10}$ | 39.176 | 39.306 | 39.134 | - |
| $[6\|6]_{10}$ | 46.986 | 46.756 | 47.430 | - |
| $[6\|8]_{10}$ | 54.796 | 54.356 | 53.930 | 54.930 |

**Table S1. Thickness analyses of the atomically controlled SRO/STO superlattices.** The total thickness of superlattice samples, obtained by XRR, PNR (data not shown for $[6|6]_{10}$ and $[6|8]_{10}$ superlattices), and STEM (*25*), consistently show the atomically designed SRO/STO superlattices with a small deviation in thickness below ~1 nm.

| Spin model | A | B | C |
|---|---|---|---|
| $M_0$ ($\mu_B$/Ru) | 0.4 | 0.3 | 0.2 |
| $M_1$ ($\mu_B$/Ru) | 0 | 0.1 | 0.2 |
| SSE | A | B | C |
| $R^+$ | 0.001118 | 0.001115 | 0.001110 |
| $R^-$ | 0.001163 | 0.001167 | 0.001218 |
| Average | 0.001141 | 0.001141 | 0.001164 |
| S. A. | 0.020479 | 0.020023 | 0.023138 |

**Table S2. SSE values of PNR spectra.** We estimated the partial FM contribution in the PNR measurement at 5 K with 0.01 T of in-plane $H$-field via comparing the sum of squared error (SSE) values. The SSE of three different spin structures (A to C spin model) is calculated by [(Simulated $M$) – (Measured $M$)]$^2$.